\documentclass[prb,aps,showpacs,superscriptaddress,epsfig,twocolumn,floatfix]{revtex4}

\usepackage[utf8]{inputenc}
\usepackage{amsmath,color}
\usepackage[english]{babel}
\usepackage[T1]{fontenc}
\usepackage{graphicx}

\begin{document}

\title{Origin of the vortex displacement field in twisted bilayer graphene}
\author{Yu. N. Gornostyrev}
\affiliation{M. N. Mikheev Institute of Metal Physics UB RAS, 620137, S. Kovalevskaya str. 18, Ekaterinburg, Russia}
\affiliation{Ural Federal University, Mira str. 19, 620002 Ekaterinburg, Russia}
\author{M. I. Katsnelson}
\affiliation{Radboud University, Institute for Molecules and Materials, NL-6525 AJ Nijmegen, The Netherlands}
\affiliation{Ural Federal University, Mira str. 19, 620002 Ekaterinburg, Russia}

\begin{abstract}
Model description of patterns of atomic displacements in twisted bilayer systems has been proposed.
The model is based on the consideration of several dislocation ensembles, employing a language that is widely
used for grain boundaries and film/substrate systems. We show that three ensembles of parallel screw dislocations 
are sufficient both to describe the rotation of the layers as a whole, and for the vortex-like displacements 
resulting from elastic relaxation. The results give a clear explanation of the observed features of the structural 
state such as vortices, accompanied by alternating stacking.
\end{abstract}


\maketitle

\section{Introduction}

Bilayer systems consisting of two layers of identical or different two-dimensional materials such as bilayer graphene (G/G), bilayer hexagonal boron nitride (BN/BN), and bilayer graphene/hexagonal boron nitride (G/BN) are the subject of great interest now as the simplest examples of ``Van der Waals heterostructures'' (for review, see Refs.\cite{GG2013,Katsnelsonbook}). Building bilayer devices involves mechanical processes such as rotation and translation of one layer with respect
to the other. This has substantial influence on the performance and quality of such devices \cite{GLi2009,JMBL2007}.
The rotation leads to structural moir\'e  patterns which directly affect the electronic properties of bilayers
\cite{WTP2005,Xue2011,Tang2013,Yang2013,Woods2014,Slotman2015,Shi2020}. A further growth of interest to the field was triggered by a recent discovery of superconductivity and metal-insulator transition in ``magic angle'' twisted bilayer graphene \cite{Cao1,Cao2}. In this paper, we focus on the structural aspects of moir\'e patterns and consider bilayer G/G. We suggest a description of the Moir\'e patterns in terms of vortices and in terms of dislocations and establish a connection between these two languages.

In two-dimensional materials, such as monolayer graphene, the term ``dislocation'' is typically used to describe pointlike (0D) defects lying within the sheet, e.g., pentagon-heptagon or square-octagon pairs \cite{Nelsonbook}; they are also used to describe grain boundaries as dislocation walls \cite{Yazyev,Akhukov,Srolovits}. Such defects are edge dislocations with line directions oriented normal to the sheet. Unlike the case of monolayers, in bilayers it is also possible to have one-dimensional (line) dislocations that lie between the two layers of a bilayer material; these dislocations do not require the generation of any topological defects within each of two sheets.

The geometry of displacement fields in bilayer Van der Waals systems has been discussed repeatedly starting
from the discovery of commensurate-incommensurate transition in G/BN system \cite{Woods2014}. The results of atomistic simulations \cite{Fasolino2014,Fasolino2015,Fasolino2016}
show that the formation of the vortex lattice  is rather typical for the picture of displacements of
the relaxed twisted bilayer systems (both G/BN and G/G).
On the other hand, the electron microscopy study \cite{JSA2013} reveals multiple stacking domains with soliton-like boundaries between them in slightly twisted bilayer graphene, the domain boundaries can be also described as
one-dimensional Frenkel-Kontorova dislocations. Wherein, a topological defect where six domains meet can be considered
as vortex in displacement field.
A similar multiple stacking domain structure was recently discussed in Ref. \cite{Bagchi2020} in the framework of a model employing a network of partial dislocation.
However, the relation between these two descriptions, in terms of vortices and in terms of dislocations, is currently unclear.

Here, based on the dislocation model, we propose a general description of the moir\'e patterns in twisted bilayer graphene it terms of twist grain boundaries in layered material. We show that both pictures (vortex network and dislocation arrays) are consistent and presented two possible ways for a qualitative description of such systems and physical interpretation of the computer simulation results.

\section{Dislocation model of a twisted bilayer system}

Moir\'e patterns are formed initially by a rigid twist of the upper layer with respect to the bottom layer; their geometry is determined by lattice type and the rotation angle \cite{Hermann2012}. If one allows atomic relaxation, that is, their shifts from these ideal geometric positions within the lattice of coincidence sites to minimize the total energy, the picture becomes more complicated and, according to simulations \cite{Fasolino2015}, vortex displacement field arises. Notably, the vortices form a regular lattice
and are separated by broad regions of almost zero displacements.

There are two canonical ways to describe conjugation in bilayer systems. The first one is the simplest picture
of coincidence site lattice (CSL) \cite{Marc74} where  one lattice just rotates and puts onto the other one without
atomic relaxation; it corresponds to the moir\'e description. In the second approach, a general description of twist boundaries
in bilayer systems can be derived on the basis of dislocation models proposed earlier for three-dimensional materials and
thin films \cite{RS1962}.
A  consideration of grain boundaries based on the concept of surface dislocations was given in the book \cite{HL1968} where
general relations between grain boundaries and geometry of dislocation arrays were discussed. In the context of graphene,
this language was used in Refs. \cite{Yazyev,Akhukov,Srolovits}.

\subsection{General geometric relationships}

At least, two arrays of parallel equidistant screw dislocations are necessary
to ensure a given relative twist of two crystallites in their conjugation plane \cite{HL1968}. In this case, certain geometric
relations must be fulfilled so that the total shear deformation in the plane of the boundary is zero.
In particular, in the case of two arrays, it is necessary to require that the dislocation axes in these arrays were perpendicular to each other.

To present correctly the geometry of conjugation of two graphene layers (and the corresponding moir\'e pattern), two dislocation
arrays are not enough. We consider more general case and represent the plastic distortion tensor
$\beta^p_{ij}$ produced by one array of dislocations in the form $\beta^p_{ij}=n_ib_j/d$, where ${\bf n}$ is the
normal to the dislocation line lying in the plane of the boundary, ${\bf b}$ is the Burgers vector of dislocation,
$d$ is the distance between neighboring dislocations in array. The plastic deformation $\varepsilon^p_{ij}$ and
rotation $\omega^p_{ij}$ are determined by the symmetric and antisymmetric parts of the tensor $\beta^p_{ij}$ \cite{HL1968}
\begin{equation}
\varepsilon^p_{ij} =\frac{n_ib_j+n_jb_i}{2d}  \  \  \  \  \  \   \omega^p_{ij} = \frac{n_ib_j-n_jb_i}{2d}
\label{eq:tensor}
\end{equation}

\begin{figure}[htbp]
\begin{center}
\includegraphics[width=3.50cm]{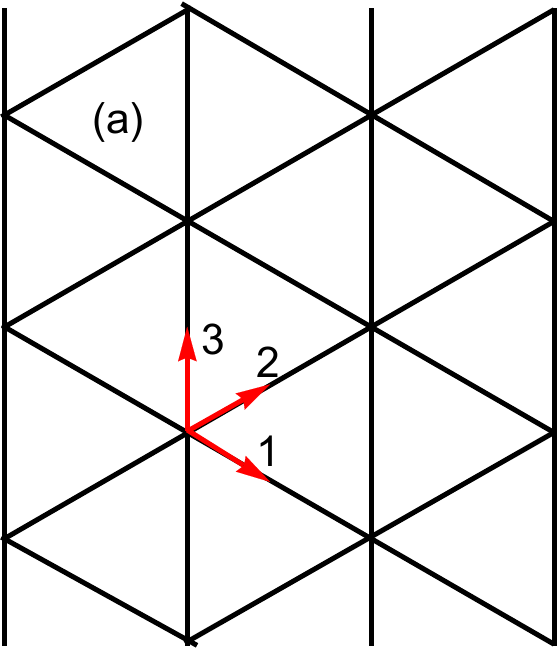} \ \ \ \ \
\includegraphics[width=3.90cm]{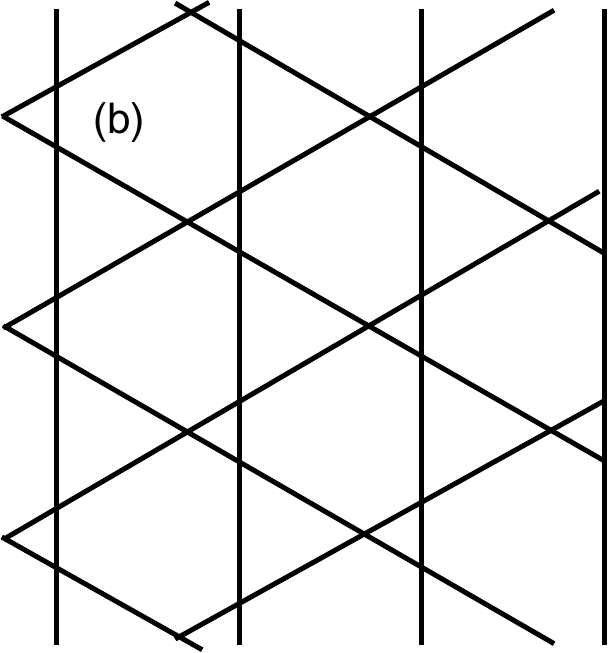}
\caption{The schematic representation of the dislocation network used to describe the twist boundary. (a) -
network of screw dislocations, (b) - reconstructed network of dislocations.
Vectors 1,2,3 indicate the directions of dislocation lines.
}
\label{network}
\end{center}
\end{figure}

To describe  the conjugation of two twisted graphene layers and taking into account the lattice trigonal symmetry, we will
use three dislocation arrays rotated with respect to each other by $\pi$/3. Assuming that the normal to the graphene layers is
$\left[ 001 \right]$ and  considering screw dislocations with Burgers vector  parallel to dislocation line,
we write
(see Fig. \ref{network}):
\begin{eqnarray}
{\bf b}_1 =b\left[ \frac{\sqrt{3}}{2}, -\frac{1}{2}, 0\right],   \  \  \   {\bf n}_1 = \left[ \frac{1}{2}, \frac{\sqrt{3}}{2},  0\right], \nonumber \\
{\bf b}_2 =b\left[ \frac{\sqrt{3}}{2}, \frac{1}{2}, 0\right],     \  \  \   {\bf n}_2 = \left[ -\frac{1}{2}, \frac{\sqrt{3}}{2},  0\right]
\label{eq:vectors12}
\end{eqnarray}
where $b$ is the module of the Burgers vector which should be equal to the elementary translation in graphene layer.
To ensure the total deformation $\varepsilon^p_{ij}$ being equal to zero, it is necessary to choose the third dislocation array
with the vector ${\bf b}_3$ orthogonal to ${\bf b}_1+{\bf b}_2$:
\begin{equation}
{\bf b}_3 =b\left[0, 1, 0\right],     \  \  \   {\bf n}_3 = \left[-1, 0,  0\right]
\label{eq:vector3}
\end{equation}
Indeed, substituting expressions (\ref{eq:vectors12}) and (\ref{eq:vector3}) into equation (\ref{eq:tensor}), we find
\begin{equation}
\varepsilon^p_{ij} =0  \  \  \  \  \  \   \omega^p_{ij} = \frac{3b}{2d}   \begin{pmatrix} 
      0 & 1 & 0 \\
      -1 &  0 & 0  \\
      0 &  0 & 0  \\
   \end{pmatrix}
\label{eq:tensor2}
\end{equation}
and the rotation vector $\omega_k = \frac12 \epsilon_{ijk}\omega_{ij}$ will be  $\omega = \dfrac{3b}{2d}\left[0,0,1\right]$.
Note that for small rotation angle $\psi \approx \omega_3=3b/(2d)$; this expression is similar to that determining the geometry
of moir\'e in the model of CSL $\psi \sim a/l$, $l$ being the distance between coincidence points,
$a$ being the elementary lattice translation.

\subsection{Relaxed displacement field within the dislocation model}

The equations (\ref{eq:tensor2}) are valid on the average for the whole sheet. In fact, the displacements are
non-uniformly distributed and concentrated near the dislocation lines. As discussed above, to describe correctly
the displacement field in the case of twisted bilayer graphene {\it three} arrays of dislocations are necessary. We believe
that the Frenkel-Kontorova model \cite{Braun} gives a qualitatively correct description of the displacement field created by
surface dislocations. Assuming that the energy relief of substrate may have an additional minimum \cite{Srolovits2,Savini}
and dislocations can split into the partial ones \cite{HL1968,Braun}, the
screw component of the displacement field for one family can be represented as
\begin{eqnarray}
u_s(x) =\frac{b}{\pi} \sum_i \left[\arctan\left(\exp\left(\frac{x-x^0_i-\delta/2}{\xi}\right)\right) \right. \nonumber \\
+ \left. \arctan\left(\exp\left(\frac{x-x^0_i+\delta/2}{\xi}\right)\right)\right]
\label{eq:displ1}
\end{eqnarray}
where $x^0_i$ correspond to position of the center of dislocation line, $\xi$ is core width and $\delta \sim \mu b/\gamma$ is
distance splitting between partial dislocations, $\mu$ is shear modulus and $\gamma$ is stacking fault energy.

The splitting of the dislocation on hexagonal lattice results in formation of stacking fault
which is accompanied also by the appearance of an edge components of partial dislocations \cite{Bagchi2020},
${\bf b}=({\bf b}/2+{\bf b}_e)+({\bf b}/2-{\bf b}_e)$. In this case, the edge
component of displacement field can be written as
\begin{eqnarray}
u_e(x) =\frac{b_e}{\pi} \sum_i \left[\arctan\left(\exp\left(\frac{x-x^0_i-\delta/2}{\xi}\right)\right) \right. \nonumber \\
- \left. \arctan\left(\exp\left(\frac{x-x^0_i+\delta/2}{\xi}\right)\right)\right]
\label{eq:displ2}
\end{eqnarray}

\begin{figure}[htbp]
\begin{center}
\includegraphics[width=6.80cm]{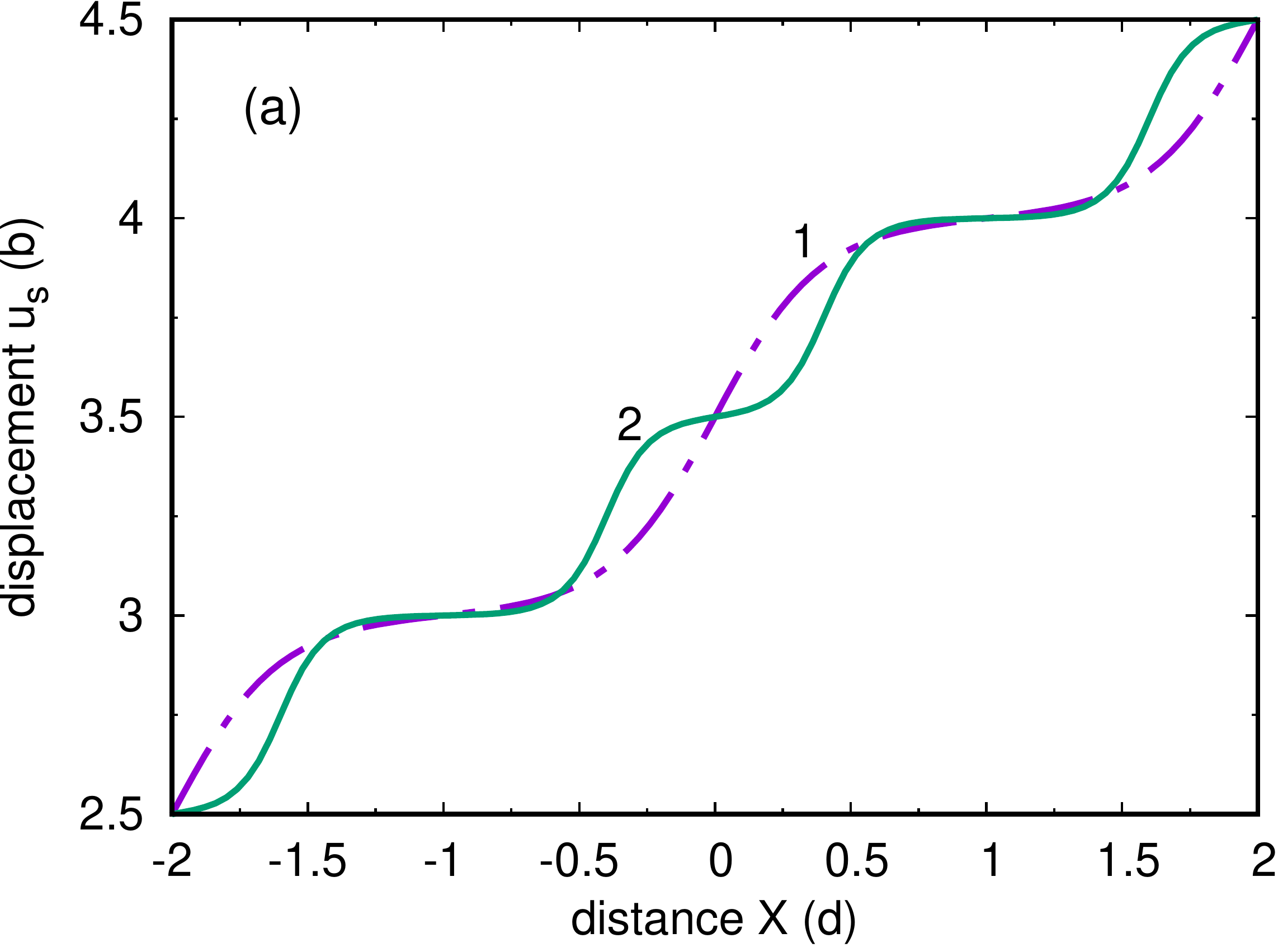} \ \ \ \
\includegraphics[width=6.80cm]{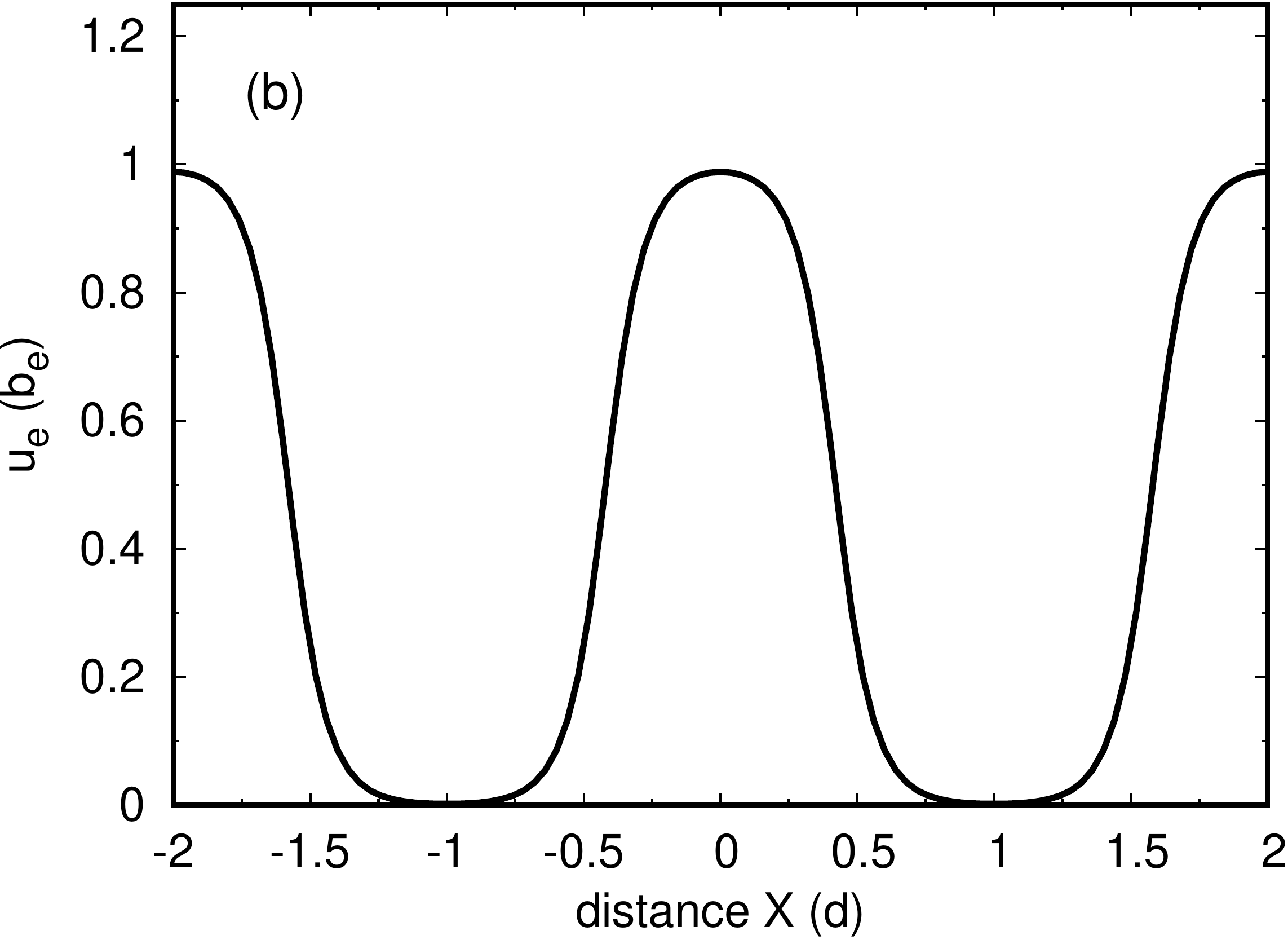}
\caption{Screw component (a) of displacements produced by one dislocation array in the
case of non-split ($\xi=0.2$, $\delta=0.2$, curve 1) and split ($\xi=0.1$, $\delta=0.8$, curve 2)
dislocation cores. Edge component (b) of the displacements is produced by one array of
partial dislocation ($\xi=0.1$, $\delta=0.8$).  Distances and parameters  $\xi$, $\delta$ are given in units of $d$.}
\label{u(x)}
\end{center}
\end{figure}

Fig. \ref{u(x)}a displays the dependence $u_s(x)$ for the cases of narrow and wide (split) dislocations.
In  the case of narrow dislocations the displacements are concentrated in the dislocation core and include both plastic
and elastic parts.
When the width of the dislocation core $\xi$ increases, the dependence $u_s(x)$ becomes close to
linear $u \approx u^p =bx/d$ and  represents a pure plastic shear.
Fig. \ref{u(x)}b shows the edge component of displacements in case of split dislocation.

Following the discussion in the previous section, we represent the total displacement field in twisted graphene
layers as a superposition of three dislocations arrays. Subtracting the plastic part, we write the elastic displacements
produced by screw dislocations in the form
\begin{equation}
\begin{aligned}
{\bf u}^{el}({\bf r}) & =   \sum_{k=1}^3 \frac {{\bf b}_k}{\pi}\sum_{i=-m}^m
\arctan\left(\exp\left(\frac{{\bf r n}_k-x^{k}_i-\delta/2}{\xi}\right)\right) \\
                         & +\arctan\left(\exp\left(\frac{{\bf r n}_k-x^{k}_i+\delta/2}{\xi}\right)\right) \\
                         & -  \left(mb+\frac {b}{L}{\bf r n}_k \right)
\label{eq:displt}
\end{aligned}
\end{equation}
where ${\bf n}_k =[001] \times {\bf b}_k $ is the normal to the dislocation line of the k-th  array.
In addition, in the case of split dislocations, there is a displacement field created by arrays of edge
partial dislocations
\begin{equation}
\begin{aligned}
{\bf u}_{e}({\bf r}) & =  \sum_{k=1}^3 \frac{{\bf b}^p_k}{\pi} \sum_{i=-m}^m \left[
\arctan\left(\exp\left(\frac{{\bf r n}_k-x^{k}_i+\delta/2}{\xi}\right)\right)\right. \\
                         & -\left. \arctan\left(\exp\left(\frac{{\bf r n}_k-x^{k}_i-\delta/2}{\xi}\right)\right)\right]
\label{eq:displt2}
\end{aligned}
\end{equation}

\begin{figure*}[thbp]
\begin{center}
\includegraphics[width=7.0cm]{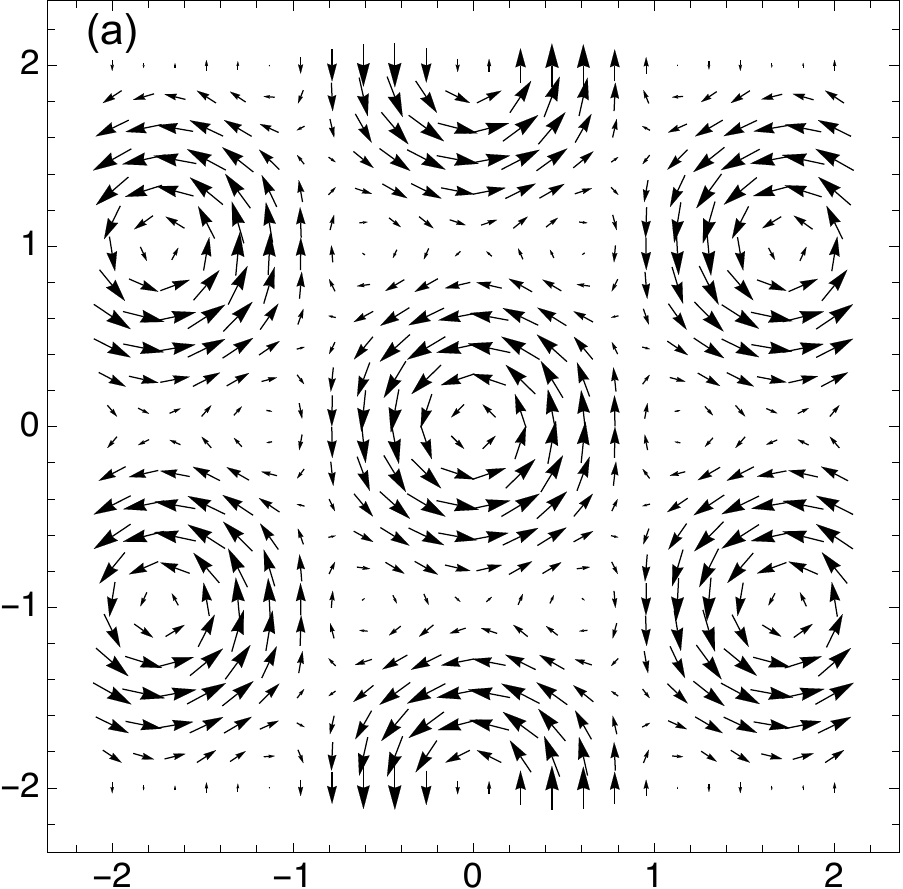} \ \ \ \ \ \
\includegraphics[width=7.0cm]{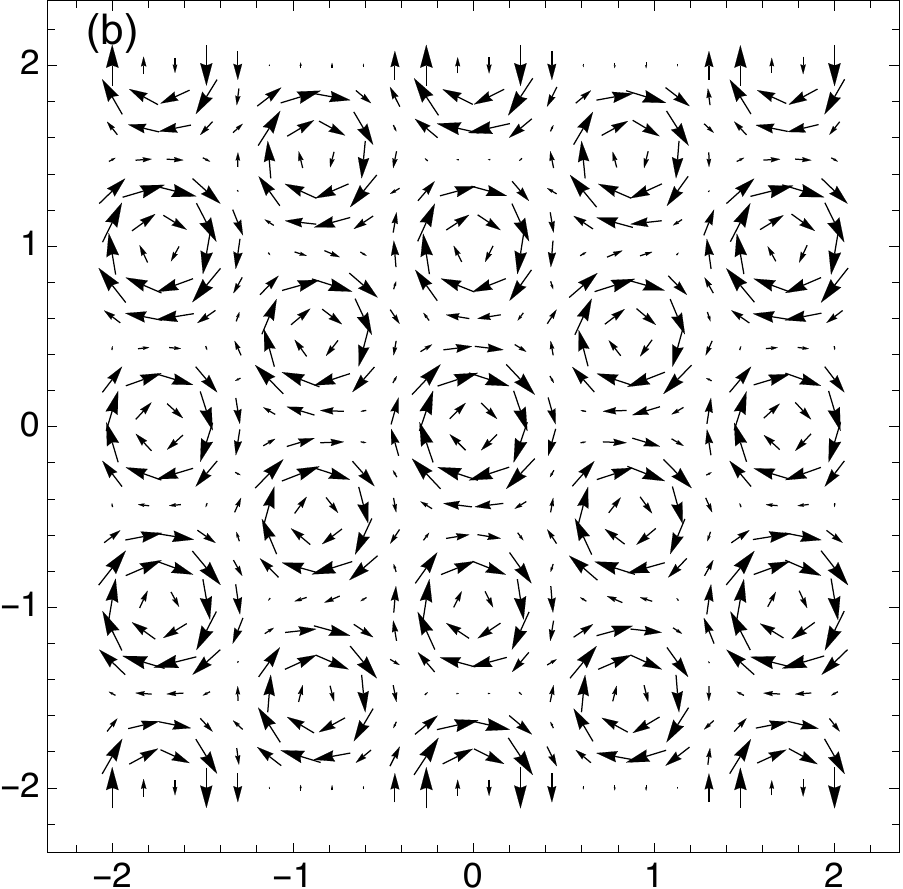} \\
\includegraphics[width=7.0cm]{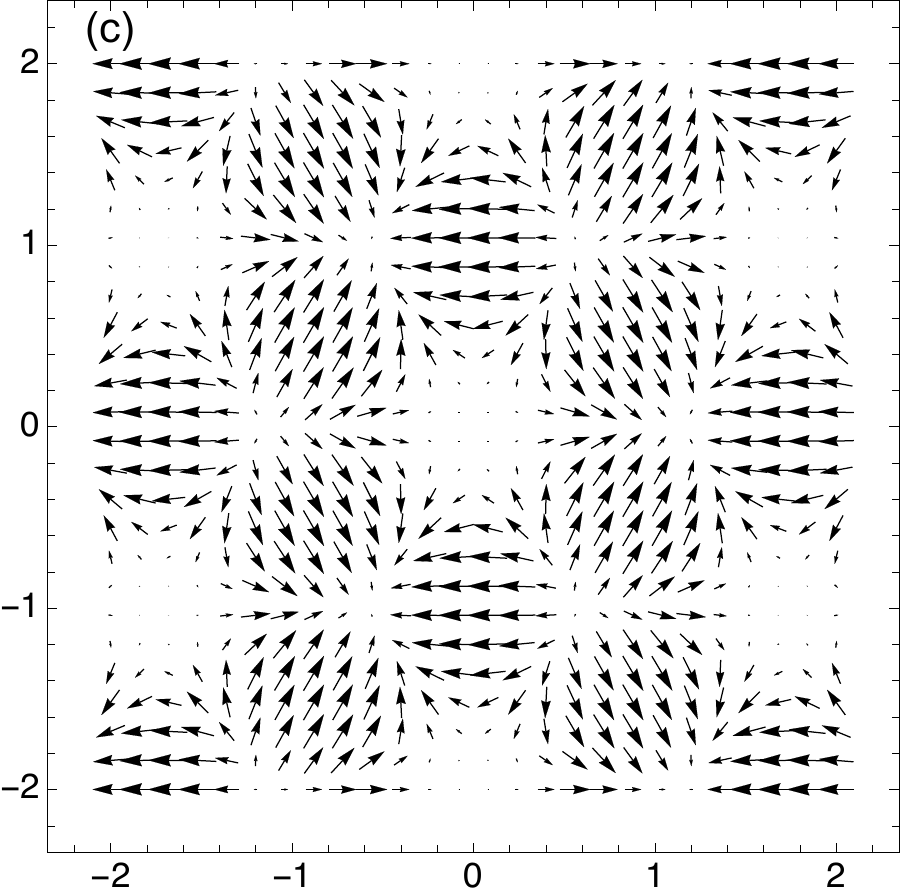} \ \ \ \ \ \
\includegraphics[width=7.0cm]{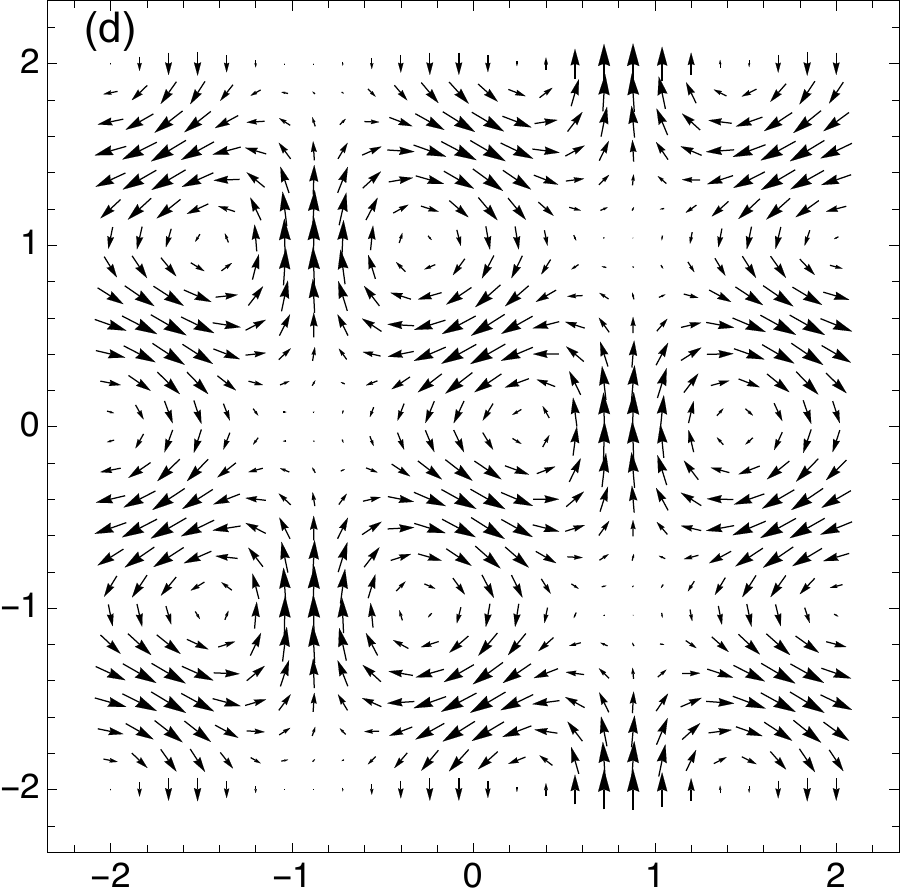}
\caption{Displacement ${\bf u}_{el}({\bf r})$ shown as a vector field for narrow (a,d), split (b,c) dislocation cores
($\delta =0.4d$), and for the reconstructed dislocation network (d). (c) - edge component of displacement field ${\bf u}_{e}({\bf r})$ in
the case of split dislocation. $\xi$ is equal 0.05$d$ in cases (a,b,c) and $\xi$ is equal 0.15$d$ in case (d).
}
\label{vecfield}
\end{center}
\end{figure*}

The vector fields described by equation (\ref{eq:displt}) are shown in Fig. \ref{vecfield} for the cases of narrow
(non-split) and split dislocation cores. As one can see from Fig. \ref{vecfield}a,b, the screw component of the
displacement field producing by elastic relaxation (\ref{eq:displt}) forms a vortex lattice.
However, the geometry of displacements is quite different for the cases under consideration.
in particular, dislocation splitting results in a decrease of the period of vortex lattice in two times;
the magnitude of displacement becomes essentially smaller (Fig. \ref{vecfield}b).
Distribution of the corresponding edge components of displacement field ${\bf u}_{edge}({\bf r})$ is shown in
 Fig. \ref{vecfield}c. Alternating domains of almost constant displacements correspond to different types
 of the stacking faults.

\begin{figure*}[htbp]
\begin{center}
\includegraphics[width=6.90cm]{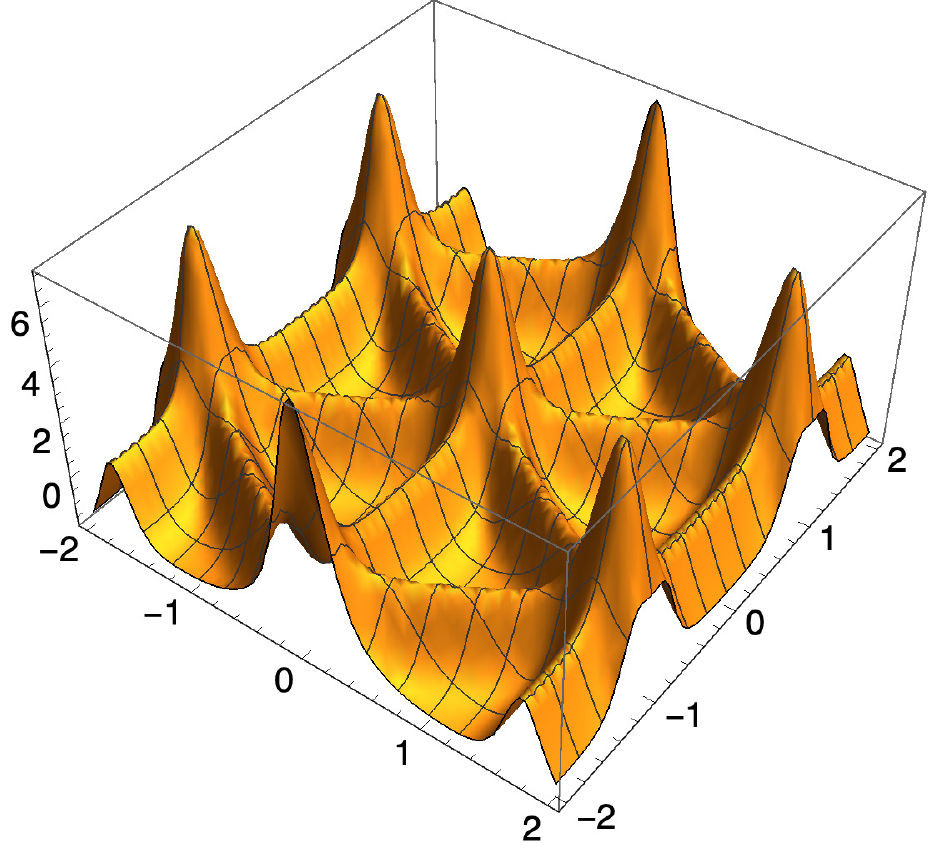} \ \ \ \
\includegraphics[width=6.90cm]{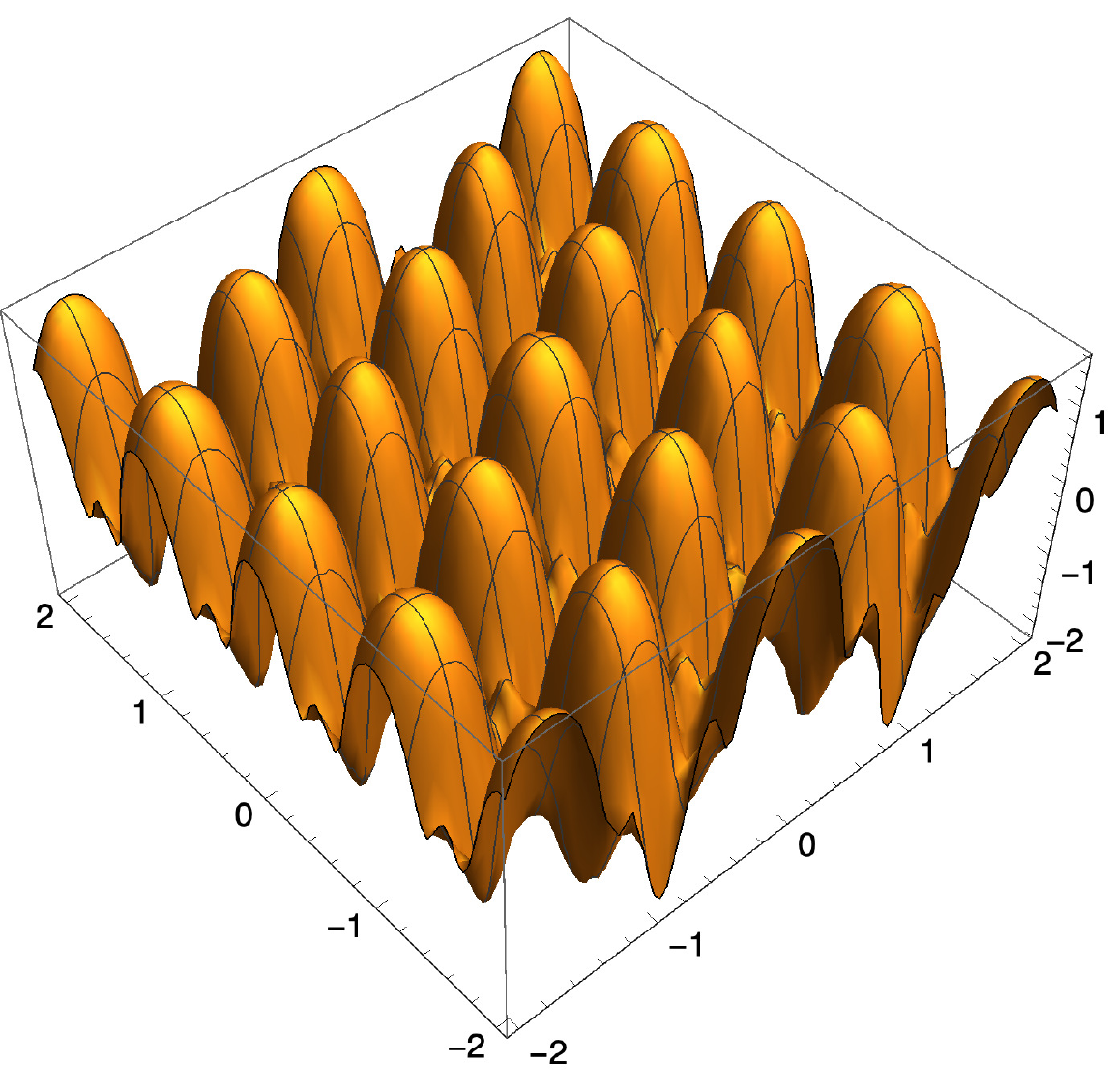} 
\caption{Distribution of z component of rotation vector ${\boldsymbol \omega}$
for narrow (left) and split (right) dislocation cores.
The corresponding displacement fields are shown in Fig. \ref{vecfield} (a) and (b).}
\label{rot}
\end{center}
\end{figure*}

The picture of elastic displacement is rather similar to that obtained in atomistic simulations (see Ref's \cite{Fasolino2015,Savini}),
which indicates a semiquantitative correctness of the description of atomic relaxation effects in twisted graphene bilayers
within our simple dislocation model.

It is worthwhile to note that in the center of vortex situated in the crossing of dislocation lines in Fig. \ref{network}
the relative displacement of the layers is equal to the half of elementary translation which results in formation of a stacking fault.
To visualize this stacking fault, one needs to pass from elastic deformations ${\bf u}_{el}({\bf r})$ to the relative displacements
between the layers described by equation (\ref{eq:displ1})). In this case the more energetically favorable configuration is the one
where one of the dislocation families is shifted from the symmetry position (Fig. \ref{network}b) by the value $\delta l$ in the direction
normal to dislocation lines.

The distribution of the elastic displacements after such a reconstruction of the dislocation network is shown
in Fig. \ref{vecfield}d. One can see that the reconstruction of the dislocation network results in a drastic change of the displacement
field ${\bf u}_{el}({\bf r})$ (c.f. Fig. \ref{vecfield} a and  d).  The regions with almost zero displacements are surrounded by six triangular regions with the largest displacements at their borders.
According to Ref. \cite{Bagchi2020} the structure of conjugation of bilayer graphene can be described in terms of
network of partial dislocations a/3<1\={1}00> separating the regions of AB and BA stacking. The picture presented in
Fig. \ref{vecfield}d agrees qualitatively with that discussed in Ref. \cite{Bagchi2020}.

The other quantity characterizing peculiarities of the displacement field is the distribution
of the rotation vector ${\boldsymbol \omega}(\bf r)= {\nabla \times \bf u}_{el}(\bf r)$ presented in Fig. \ref{rot}. As one can
see from this figure, in the case of narrow core (Fig. \ref{rot}a), the dislocation lines are clearly visible in the distribution of
the rotation vector and their intersections correspond to the centers of vortices. At the same time, the only lattice of vortices remains
visible in the case of split dislocations.
Thus, depending on the representation used, the conjugation of layers can
be described either in terms of vortices or dislocations.
Wherein, the vortex displacement field originates naturally from elastic relaxation of atomic positions.
Note that the magnitude of rotation vector $\omega$ is close to zero for the edge component of displacements.

The other quantity which is actively discussed now \cite{pseudomag} is the distribution of pseudomagnetic field (PMF)
\cite{Katsnelsonbook,physrep} given by the equations

\begin{equation}
H_{PMF}= \frac{dA_y}{dx}-\frac{dA_x}{dy}
\label{pmf}
\end{equation}
where vector potential  is expressed (with the accuracy of some constant multiplier) via deformations as
\begin{equation}
A_{x}= u_{xx}-u_{yy}, \ \ \ \ \\ A_{x}= -2u_{xy}, \nonumber
\label{vpot}
\end{equation}
\begin{eqnarray}
u_{xx}=\frac{du_x}{dx}+\frac{1}{2}  \left(\frac{du_y}{dx} \right)^2 +\frac{1}{2}  \left( \frac{du_y}{dx}\right)^2, \nonumber \\
u_{yy}=\frac{du_y}{dy}+\frac{1}{2}  \left( \frac{du_y}{dy}\right)^2 +\frac{1}{2}  \left( \frac{du_x}{dy}\right)^2, \nonumber \\
u_{xy}=\frac{1}{2} \left(  \frac{du_x}{dy}+ \frac{du_y}{dx} + \frac{du_x}{dy}\frac{du_x}{dx} + \frac{du_y}{dy}\frac{du_y}{dx} \right)
\label{def}
\end{eqnarray}

\begin{figure}[htbp]
\begin{center}
\includegraphics[height=7.0cm]{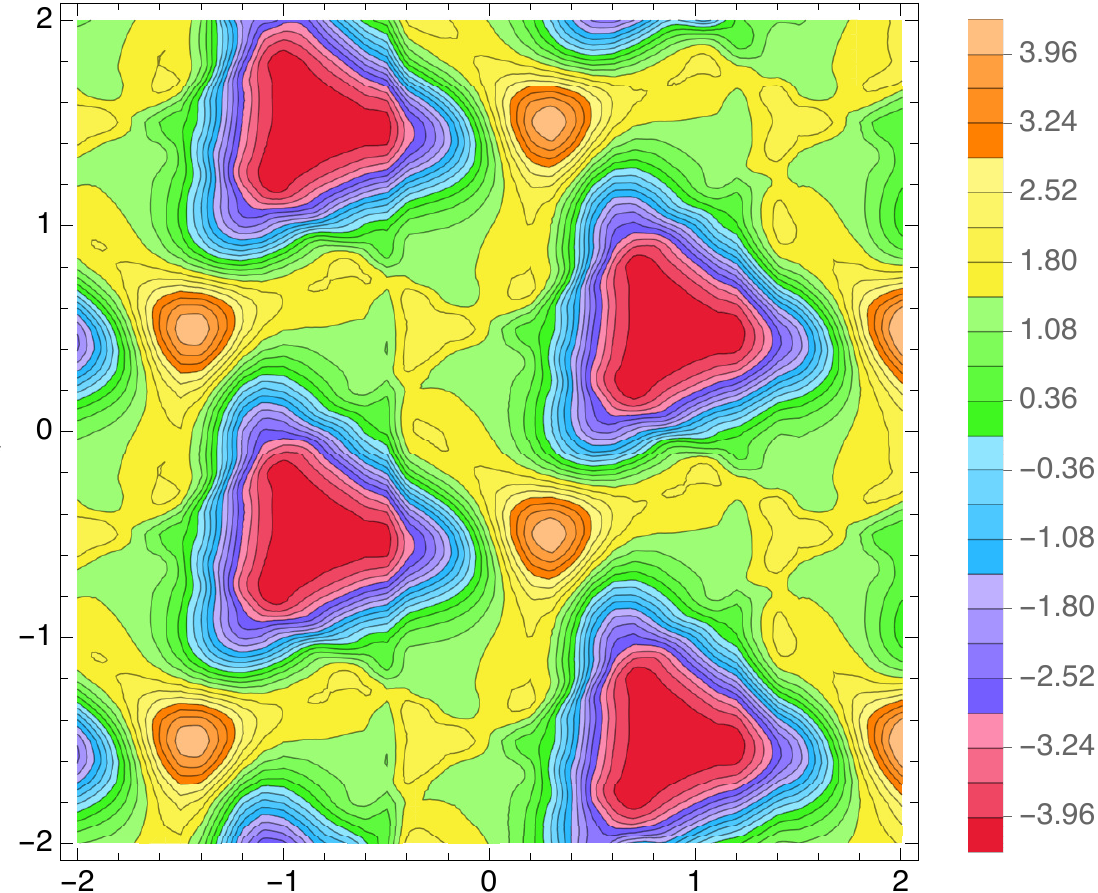}
\caption{Distribution of PMF  calculated for reconstructed dislocation network (Fig. \ref{vecfield}d) by using
Eq's (\ref{pmf})-(\ref{def}) }
\label{dist}
\end{center}
\end{figure}

The distribution of PMF calculated from equations (\ref{pmf})-(\ref{def}) for the reconstructed
dislocation network with the  displacement field from Fig. \ref{vecfield}d is shown in Fig. \ref{dist}. Pink regions
(with negative PMF) correspond to the domains of small displacements in Fig. \ref{vecfield}d;
quasicircular yellow regions with positive PMF are situated in between. The presented picture is characteristic of the
reconstructed dislocation network and will be much less regular for the other distributions of displacement fields
considered here.

\section{Discussion and conclusions}

The model developed here allows us to explain naturally some qualitative peculiarities of the moir\'e patterns in
bilayer systems. In particular, the rotation angle and location of the coincidence site of the moir\'e  patterns are related to
Burgers vector and the distance between dislocation lines (see Section IIa).
We show that the plastic part of the displacement field provides rotation of one layer with respect to the other one as a whole.
At the same time, the distribution of elastic displacements is quite complicated and vortices are the most typical element;
the observed picture is in a qualitative agreement with both experiment \cite{JSA2013} and the results of atomistic simulations \cite{Fasolino2014,Fasolino2015,Fasolino2016,Bagchi2020}.

Although, by construction, the displacements $\bf u^{el}$ are equal to zero at a point located in the middle between two
vortices, the derivative $d u^{el}_i/d x_j \neq 0$. Note that the values $d u^{el}_i/d x_j $ depend on distance between the
dislocations in a given array and the width of their core. For narrow dislocations located far enough from each other
$d u^{el}_i/d x_j \approx 0$ in the middle between two vortices. However, a more realistic consideration of G/G case \cite{Srolovits2,Savini,Bagchi2020} 
assumes the dislocation core splitting into partial dislocation cores.
This means that the total dislocation core width is rather large,
 and the cores are situated rather close to each other. As a consequence, one should expect
a remarkable deviation of the values $d u^{el}_i/d x_j $ from zero between the vortices.
By using equation (\ref{eq:displt}) and assuming $L \gg z$ we have $d u^{el}_x/d x_y \approx 2b/\pi z \exp(-L/2z)\cosh(d/z)$.
As a result, the region between the vortices is characterized by an excess of elastic energy density
$\Delta E_{el} \sim \mu (d u^{el}_x/d x_y)^2$. This additional energy can be reduced by self-reorientation of the graphene layers
\cite{Fasolino2016} resulting in the increase of the distance between dislocations (c.f. Eq. (\ref{eq:tensor2})).

Note, that although the model does not take into account a details related to the characteristics
of chemical bonding and the formation of various types of stacking, it provides a clear vision of qualitative
structural features of twisted bilayer systems. Our results unify the language used in the physics of moir\'e patterns 
in twisted bilayer graphene and other Van der Waals heterostructures with that traditionally used at the description 
of nano- and mesostructures in solids. We suggest explicit analytical expressions for the distribution of atomic 
displacements in twisted bilayer graphene which can be used for both model theoretical studies and interpretation 
of experimental and computational results.

\section{Acknowledgements}

This work of MIK was supported by the JTC-FLAGERA Project GRANSPORT and work of YNG was financed
by the Russian Ministry of education and science (topic "Structure" A18-118020190116-6).

\end{document}